%% file: proc_jklein_bari.tex
\documentclass[sort&compress,5p,twocolumn]{elsarticle}

\usepackage[english]{babel}
\usepackage{amssymb}
\usepackage{upgreek}
\usepackage{xspace}
\usepackage{color}
\usepackage{graphicx}
\usepackage{hyperref}
\usepackage{lineno}

\journal{NIM A}

\newcommand{\pt}{$p_\perp$\xspace}

\newcommand{\pbpb}{Pb--Pb\xspace}

\begin{document}

\begin{frontmatter}

  \title{Triggering with the ALICE TRD}

  \author[pi]{Jochen~Klein}
  \ead{jklein@physi.uni-heidelberg.de}

  \author{for the ALICE collaboration}

  \address[pi]{Physikalisches Institut, University of Heidelberg,
    Philosophenweg~12, 69120~Heidelberg, Germany}

  \begin{abstract}
    We discuss how a level-1 trigger, about $8\, \upmu \mathrm{s}$
    after a hadron-hadron collision, can be derived from the
    Transition Radiation Detector (TRD) in A Large Ion Collider
    Experiment (ALICE) at the LHC. Chamber-wise track segments from
    fast on-detector reconstruction are read out with position, angle
    and electron likelihood. In the Global Tracking Unit up to
    6~tracklets from a particle traversing the detector layers are
    matched and used for the reconstruction of transverse momentum and
    electron identification. Such tracks form the basis for versatile
    and flexible trigger conditions, e.g. single high-\pt hadron,
    single high-\pt electron, di-electron ($\rm J/\Uppsi$,
    $\Upupsilon$) and at least $n$ close high-\pt tracks (jet).

    The need for low-latency on-line reconstruction poses challenges
    on the detector operation. The calibration for gain (pad-by-pad)
    and drift velocity must be applied already in the front-end
    electronics. Due to changes in pressure and gas composition an
    on-line monitoring and feedback loop for these parameters is
    required. First experiences on the performance were gathered from
    triggering in cosmic and pp runs.
  \end{abstract}


\end{frontmatter}

\section{Introduction}

The design of the ALICE TRD was driven by the requirement for very
good pion rejection in the high multiplicity environment of \pbpb
collisions and by the goal to use the information for a fast trigger
contribution. Many interesting probes, such as $\rm J/\Uppsi$,
$\Upupsilon$, open heavy-flavour, have (semi-)electronic decay
channels. For their analysis good electron identification is of
crucial importance. As these probes are rare, triggering is essential
to record a sufficient sample of events. Covering $[-0.9, 0.9]\times 2
\pi$ in $\eta$-$\varphi$, the scope of the trigger extends beyond
electron channels, e.g. to jets.

The detector is segmented into 18 super-modules in azimuth, each
comprising 5~stacks of 6~layers of tracking chambers. Each chamber
consists of a radiator, a drift volume and a multi-wire proportional
chamber with pad read-out~\cite{tdr:trd}. In this design a short drift
could be combined with the possibility of local chamber-wise
tracking. For efficient absorption of the transition radiation photons
Xe is used as counting gas, with a 15\, \% admixture of CO$_2$ as
quencher. The front-end electronics is mounted directly on the
chambers. The general setup and performance of the TRD and its physics
applications were discussed in other contributions at this
meeting~\cite{proc:trd:perf,proc:trd:phys}.

In section~\ref{sec:overview} the general concept of the TRD triggers
is described providing an overview of the steps involved. The
subsequent section explains the on-line reconstruction in more
detail. Sections~\ref{sec:sim} and~\ref{sec:perf} desribe the
simulation framework and the observed performance, respectively. In
section~\ref{sec:jet_trigger} a TRD trigger on jets is discussed.

\section{TRD-based triggering}
\label{sec:overview}

\begin{figure}[h]
  \centering
  \includegraphics[width=0.4\textwidth]{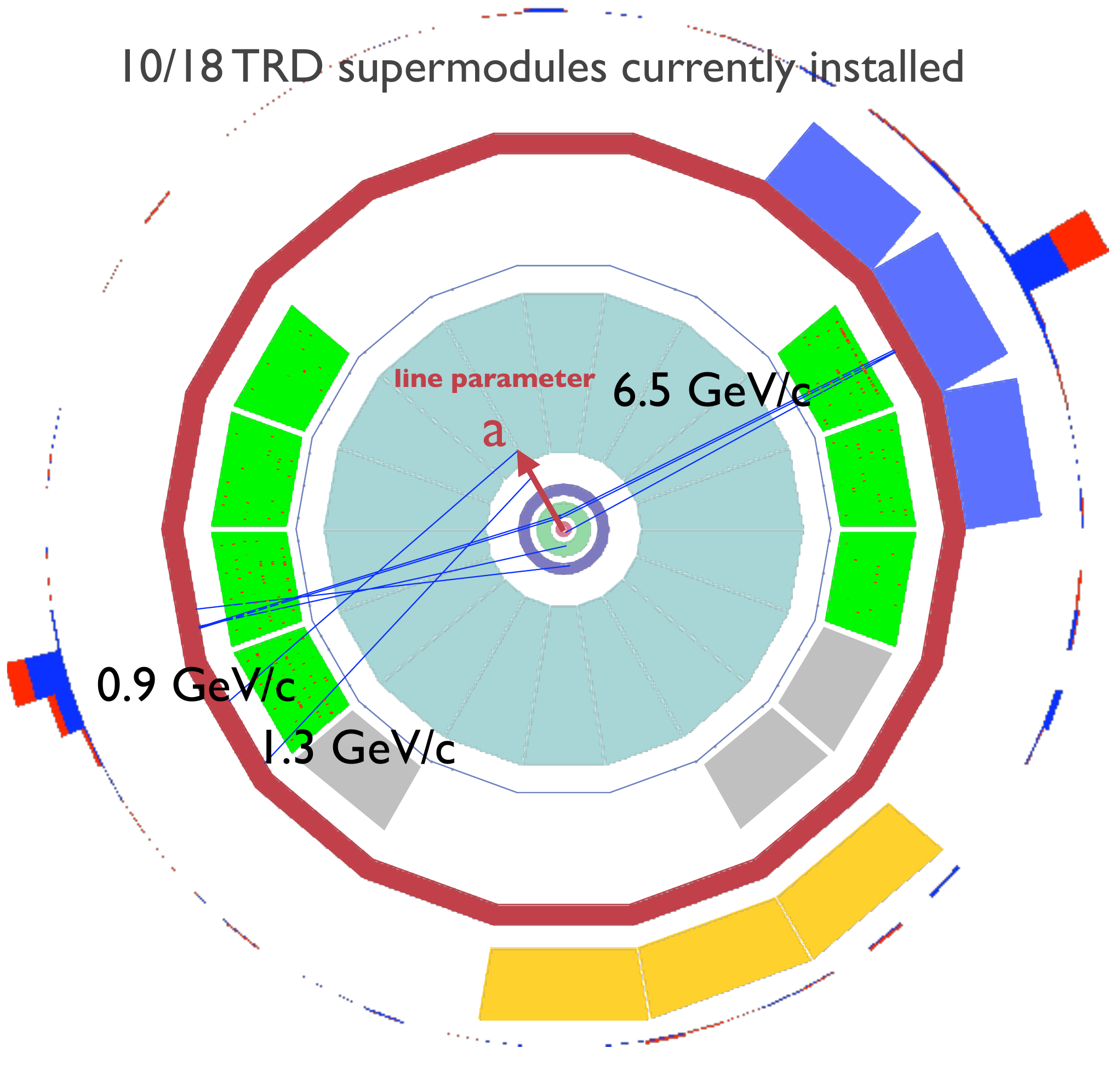}
  \caption{An event display showing two back-to-back jets of about
    $45~\mathrm{GeV}$. The blue lines show the global tracks
    reconstructed as straight line fits through the TRD tracklets. The
    line parameter $a$ is used to extract the transverse momentum.}
  \label{fig:evdisp}
\end{figure}

In ALICE three levels of hardware triggers are used. At the first
stage level-0 contributions from fast detectors, such as scintillators
and silicon pixel detectors, are evaluated by the Central Trigger
Processor (CTP) to issue a level-0 trigger. The set of triggered
detectors depends on the combination of fired trigger
inputs. $6.5~\upmu\mathrm{s}$ after a level-0 trigger the read out can
be continued or aborted depending on the contributions to the level-1
trigger. So far, the ElectroMagnetic CALorimeter (EMCAL), the
Zero-Degree Calorimeter (ZDC) and the TRD contribute at that
level. After the drift time of the time projection chamber ($\sim 90\,
\upmu\mathrm{s}$) the read-out must be accepted or rejected by a
level-2 decision. This mechanism is mainly intended for the rejection
of events with too many pile-up interactions (past-future protection)
but could also be used for more complex triggers. The trigger chain is
completed by the High-Level Trigger (HLT) implemented in a computing
cluster. It processes the data from all events which have been
accepted at level-2. For an overview of possible rates at the
different trigger stages see Table~\ref{tab:trg}.

Data from the TRD become available only after the acquisition has been
initiated by a level-0 trigger. Therefore, a high level-0 rate is
needed to sample a large number of events with the TRD triggers. They
are based on tracks which are reconstructed on-line from TRD data
only. First, local track segments, so-called tracklets, are calculated
in every chamber. Those are shipped via optical links to the Global
Tracking Unit (GTU). To achieve low latencies all tracklets must be
accepted without backpressure. The tracklet read-out is sorted such
that the matching can be performed as the data arrive. For found
tracks the transverse momentum is extracted from a straight line fit
as the offset to the primary vertex.

\begin{table}
  \centering
  \begin{tabular}{lrr}
    \multicolumn{1}{c}{\bf trigger} & \multicolumn{1}{c}{\bf time} & \multicolumn{1}{c}{\bf rate}\\ \hline\hline
    level-0 & $\sim 1.2\ \upmu\mathrm{s}$ & $\sim
    100\ \mathrm{kHz}$\\
    level-1 & $\sim 7.7\ \upmu\mathrm{s}$ & $\sim
    2.5\ \mathrm{kHz}$\\
    level-2 & $\sim 100\ \upmu\mathrm{s}$ &
    $\sim 1.5\ \mathrm{kHz}$
  \end{tabular}
  \caption{Overview of the hardware trigger stages used in ALICE. The
    three hardware levels are followed by a computing cluster forming
    the High-Level Trigger \cite{pap:alice}.}
  \label{tab:trg}
\end{table}

By using the tracks as basis for the trigger decision a variety of
signatures can be used. A trigger on a single high-\pt particle, on
top of a level-0 condition based on TOF, has been used for cosmics
data taking. In this way a very pure sample of tracks could be
provided for the analysis of very high momentum cosmic
muons~\cite{proc:trd:tr}. Asking for several tracks above a \pt
threshold in a small $\eta-\varphi$~area allows to trigger on
jets. The selection of identified electrons above a \pt threshold
allows to enhance events with semi-leptonic decays of heavy flavour
mesons. The di-electron signature shall be used to select events with
electronic decays of $\rm J/\Uppsi$ or $\Upupsilon$.

\section{On-line reconstruction}
\label{sec:reco}

\subsection{Local tracking}

\begin{figure}
  \centering
  \includegraphics[width=0.4\textwidth]{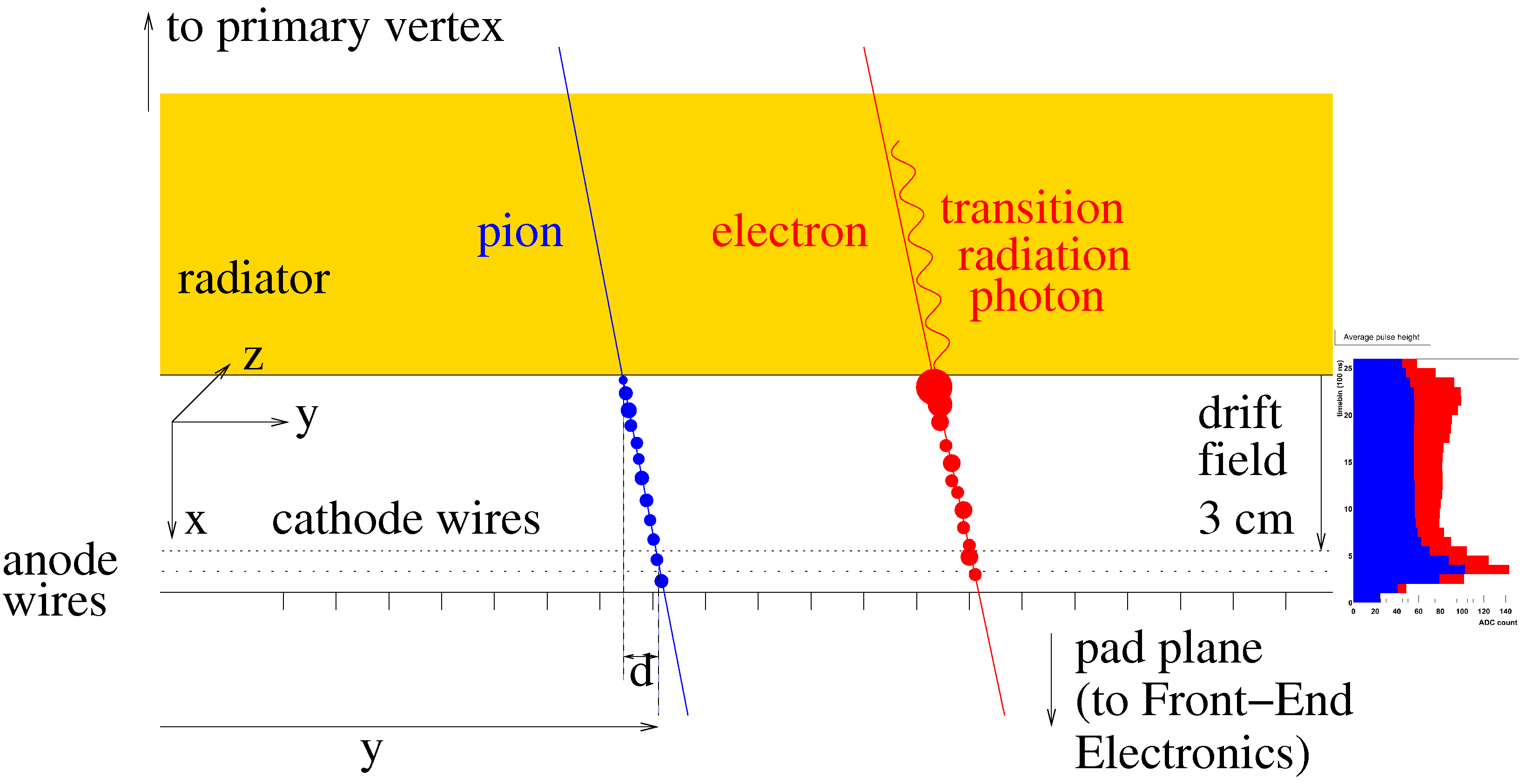}\\[.4cm]
  \includegraphics[width=0.4\textwidth]{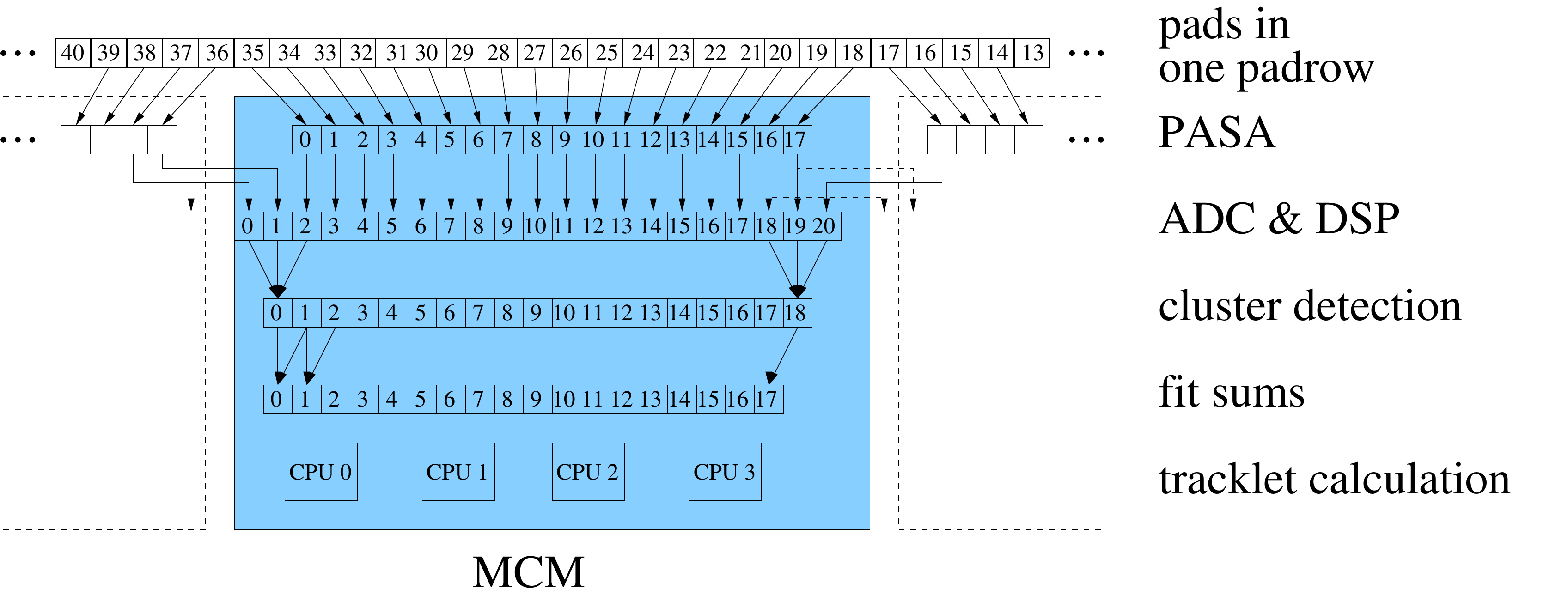}
  \caption{Top: Cross section of a TRD chamber. Depending on its
    Lorentz $\gamma$ a charged particle traversing the radiator can emit
    transition radiation which is absorbed within the drift
    region. Together with the electrons from ionisation of the gas the
    signal is detected in a MWPC. The pads are read out by front-end
    electronics mounted on the back of the chamber. Bottom: The
    processing of the pad singals. The charge-sensitive PASA feeds the
    ADCs in the TRAP. The edge channels are shared among adjacent
    chips. A preprocessor prepares the data for the final tracklet
    fit by the CPUs.}
  \label{fig:part_trav}
\end{figure}

When a charged particle traverses a TRD chamber (see
Fig.~\ref{fig:part_trav}) it deposits energy by ionization of the gas
in the active volume of the detector. Furthermore, highly relativistic
particles with a Lorentz factor $\gamma \gtrsim 1000$ can emit
transition radiation in the radiator material consisting of a
fibre/foam sandwich~\cite{tdr:trd}. Because of the high
photoabsorption cross section in Xe such a photon with X-ray energy is
absorbed most likely close to the entrance to the drift volume. Except
for very energetic particles, e.g. cosmic muons, the characteristic
additional energy deposit at the end of the drift time can be
exploited for the identification of electrons.

The electrons from ionization of the gas drift towards an
amplification region which is separated from the drift volume by a
cathode wire grid. Induced signals are read out from a cathode
pad-plane. Its granularity is fine in the tranverse direction ($\sim
1\, \mathrm{cm}$), while it is coarse along the beam direction ($\sim
10\, \mathrm{cm}$). Therefore, the $y$-position (bending direction)
can be determined accurately for individual clusters. To recover the
$z$-position during off-line tracking the pads are tilted by
$\pm2^\circ$~degrees with the sign alternating from one layer to the
next.

The signal from the pads is fed to the front-end electronics mounted
on the backside of the pad plane. Groups of 18~channels are connected
to a Multi-Chip Module (MCM) with two ASICs. The output of the
charge-sensitive PreAmplifier and Shaping Amplifier (PASA) is fed to
the TRAcklet Processor (TRAP). To avoid inefficiencies for the on-line
tracking the edge channels are shared between adjacent TRAPs as shown
in Fig.~\ref{fig:part_trav}. The signals are digitized by ADCs with a
sampling frequency of 10~MHz. With a drift velocity around $1.5\,
\frac{\mathrm{cm}}{\upmu\mathrm{s}}$ the full signal extends over
20~time bins. Typically, 24~time bins are read out to cover both the
rising and falling edge of the signal, which are needed e.g. for the
drift velocity calibration.

The data pass through a chain of digital filters. A pedestal filter
subtracts the channel-specific baseline as obtained over a long
sampling period. A common baseline is added again to allow for the
detection of undershoots created in later filter stages. Gain
variations can be corrected for each of the 1.3~million channels by
applying a correction:
\begin{equation}
  O_n(t) = \gamma_n \cdot I_n(t) + \alpha_n
\end{equation}
where $n$ denotes the channel number. Because the multiplicative
correction~$\gamma_n$ also affects the common baseline an additive
correction~$\alpha_n$ is needed to readjust the baseline. The
so-called gain tables with the correction values for all channels are
obtained from calibration runs with meta-stable Krypton added to the
gas system~\cite{proc:trd:kr} and are loaded during the initialization
of the front-end electronics. In a tail cancellation filter the ion
tails are suppressed by subtraction of two exponentials:
\begin{equation}
  S(t) = 1_{(t \geq 0)} \cdot \left( \alpha_\mathrm{L}
  \lambda_\mathrm{L}^t + (1 - \alpha_\mathrm{L}) \lambda_\mathrm{S}
  \lambda_\mathrm{S}^t \right) \mbox{ .}
\end{equation}
Both the decay rate and the relative contribution are configurable as
$\lambda_\mathrm{L}$, $\lambda_\mathrm{S}$, and
$\alpha_\mathrm{L}$~\cite{pap:trd:trap}.

\begin{figure}
  \centering
  \includegraphics[width=0.4\textwidth]{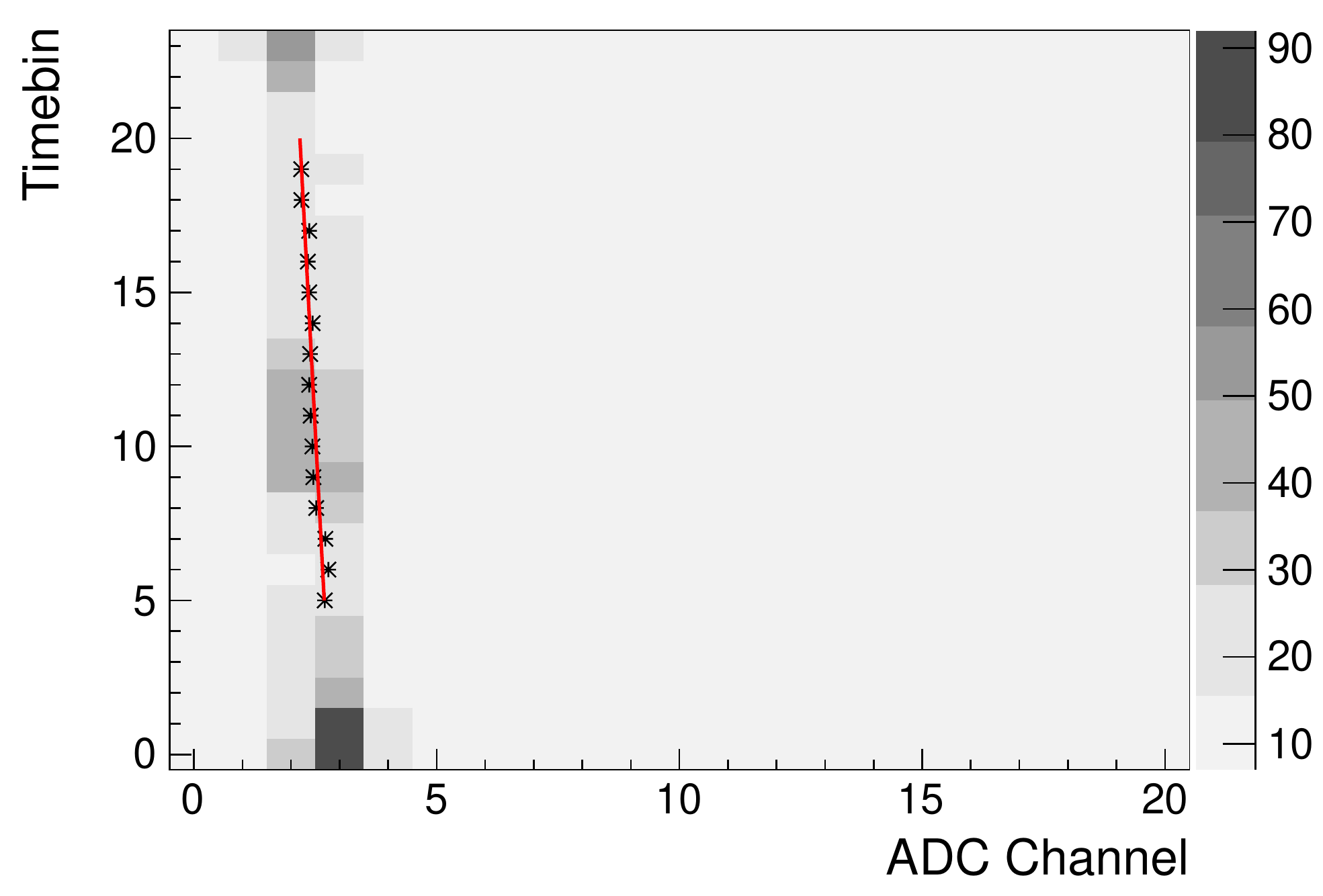}
  \caption{Processing of the ADC data within one MCM. The digitized
    values are searched time bin wise for clusters, i.e. charge in
    three adjacent channels. The calculation of the hit position
    (asterisks) takes into account the pad response function. If a
    sufficient number of clusters is found in two adjacent channels a
    tracklet fit is calculated as straight line.}
  \label{fig:mcm}
\end{figure}

The filtered data are searched timebin-wise for clusters (see cluster
detection in Fig.~\ref{fig:part_trav}). To find a cluster the charge
on three adjacent pads must exceed a configurable threshold. In
addition, the central pad must have the highest signal. For every
found cluster a position is calculated as the centre of gravity of the
charges to which a correction from a configurable look-up table is
added. This look-up table is filled with values obtained from the
known pad response functions. The data available at that stage is
shown in Fig.~\ref{fig:mcm}.

Later, the clusters shall be used for linear fitting. Therefore, the
needed sums of $X$, $X^2$, $Y$, $Y^2$ and, $XY$ are accumulated during
the processing for all channels in parallel. Each cluster is stored
according to its central channel. $X$ denotes the time bin and $Y$ the
position in units of pad widths.

To keep the noise level low the CPUs are only started after the drift
time and, thus, after the sampling. For each pair of adjacent channels
the numbers of hits is counted over all time bins in a configurable
range. If a minimum number of hits is exceeded the tracklet
calculation is performed. Up to four candidates can be processed by
the CPUs. Position and slope are calculated from the accumulated fit
sums.

Initially, the slope is calculated per timebin. To obtain the
deflection $d$ (see Fig.~\ref{fig:part_trav}) it is multiplied by the
drift velocity and the drift length of $3\, \mathrm{cm}$. In addition,
it is corrected for the Lorentz drift and the pad tilt. A cut on the
deflection is applied to reject tracklets from low \pt tracks. If the
cut was passed the final information of a tracklet is packed into a
32-bit word and shipped over the optical read out tree to the Global
Tracking Unit.

\begin{figure}
  \resizebox{0.4\textwidth}{!}{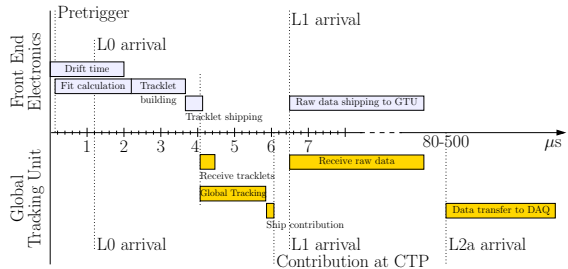}
  \caption{Timing requirements for the on-line reconstruction. The
    electronics is woken up by a pretrigger signal shortly after the
    interaction. Only for a level-0 accept the tracklet calculation
    proceeds. The raw data readout only starts after a level-1
    trigger.}
  \label{fig:timing}
\end{figure}

The temporal sequence of the various steps is shown in
Fig.~\ref{fig:timing}. After receival of a pretrigger the full drift
time can be sampled because of pipeline stages in the digital
filtering. The fit sums are calculated during the drift time, but the
CPUs only start after the end of the drift time. The tracklets are
processed in the GTU while arriving. If a given signature was found
the trigger contribution is shipped to the CTP. The raw data readout
only starts after a level-1 trigger was received.

\subsection{Global Tracking}

The global tracking is implemented in an array of
FPGAs~\cite{pap:gtu}. Upon arrival at the Global Tracking Unit (GTU)
the tracklets are grouped according to their $z$-position. The
subsequent matching stage only considers combinations of tracklets
that are consistent with a track pointing to the primary vertex in the
longitudinal direction. The tracklets are projected to a reference
plane. If tracklets from at least 4 layers fall into a close region
they form a track. For sets of matched tracklets the \pt
reconstruction is performed by calculating a straight line fit through
the tracklet positions in the transverse direction. The line
parameter~$a$ is translated into the transverse momentum. An overview
of the procedure is given in Fig.~\ref{fig:evdisp}.

\subsection{Electron identification}

\begin{figure}
  \centering
  \includegraphics[width=0.45\textwidth]{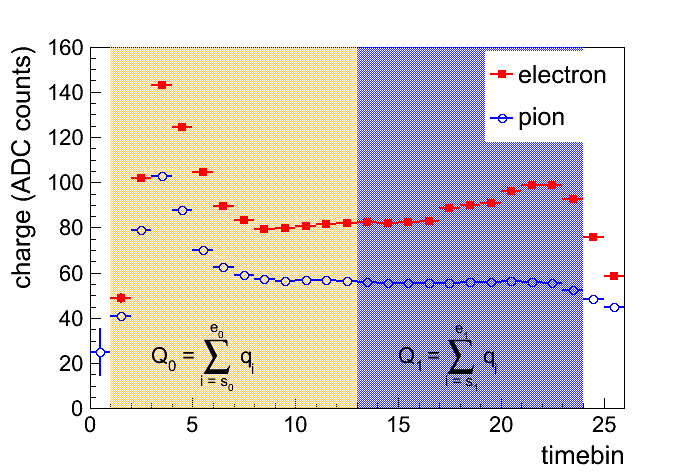}
  \caption{The on-line electron identification is based on the charge
    accumulated in two configurable time windows. The tracklet PID for
    $Q_0, Q_1$ is then looked up from a table loaded to the TRAP
    memory. It is also possible to use only one window and use the
    total charge for the PID.}
  \label{fig:pid}
\end{figure}

Electron likelihoods are carried along through the tracking chain. For
each tracklet the charge of found clusters is summed within two
configurable time windows during local tracking as illustrated in
Fig.~\ref{fig:pid}. Then, a look-up table is used to translate these
charges into an electron likelihood. For the global particle
identification used for track selection in trigger conditions, the
likelihoods from the tracklets are averaged in the Global Tracking
Unit.

For the extraction of reference spectra from real data samples of
topologically identified electrons and pions, e.g. from $\rm \upgamma
\rightarrow e^+e^-$ or $\rm K^0 \rightarrow \uppi^+\uppi^-$, are
used. A one-dimensional look-up table using the total charge of a
tracklet is used at the moment to reduce the required statistics for
the reference data extraction.

\begin{figure}
  \centering
  \includegraphics[width=0.4\textwidth]{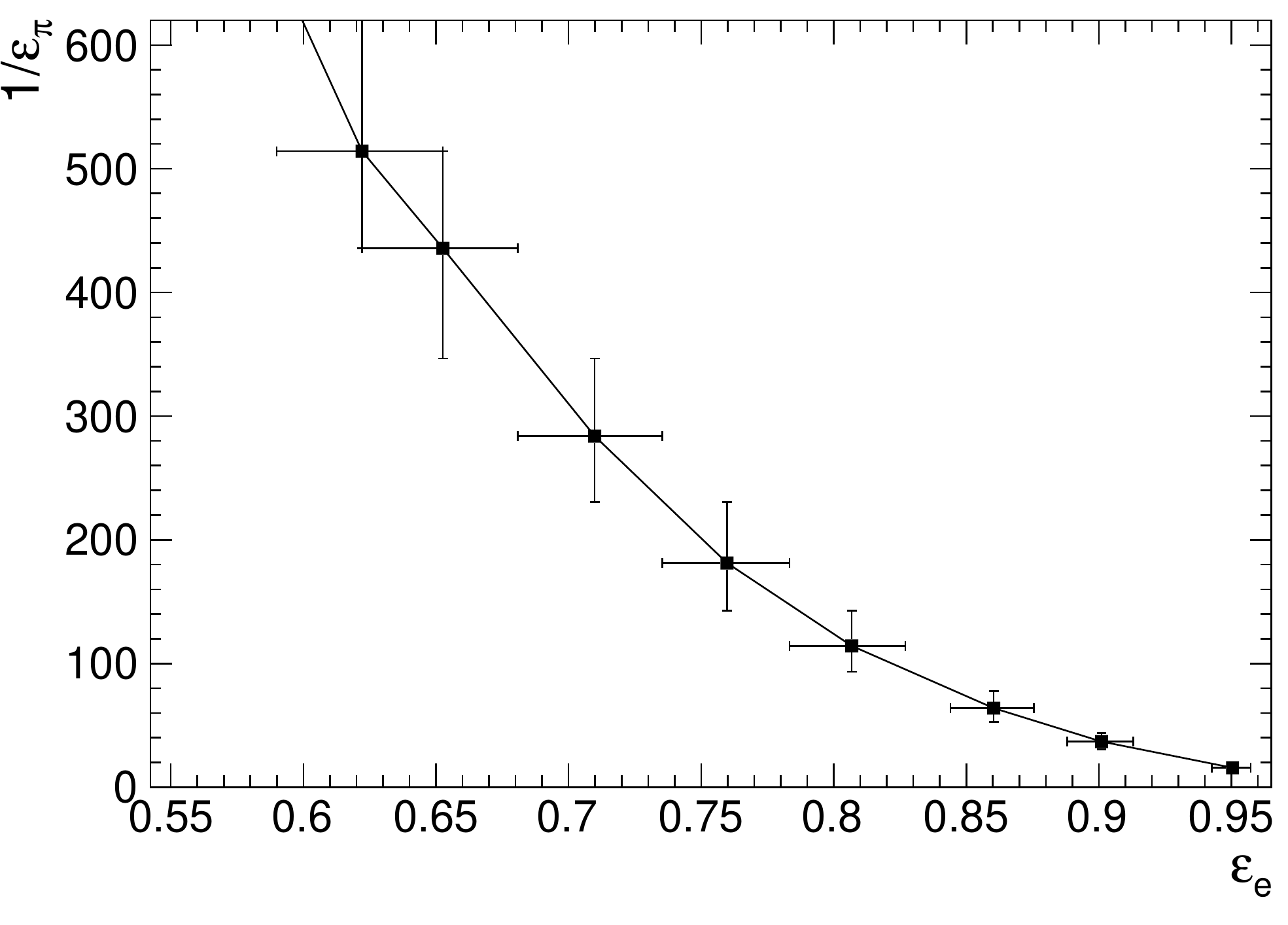}
  \caption{Pion rejection in Monte-Carlo simulation. The likelihood
    threshold to identify a particle as an electron is chosen such
    that a given electron efficiency $\epsilon_e$ is achieved. The
    resulting pion rejection $1/\epsilon_\pi$ can then be
    extracted (adapted from \cite{th:hess}).
  }
  \label{fig:pion_rej_mc}
\end{figure}

In Monte-Carlo simulations the look-up table can be generated from the
signals observed for particles of known species. Then, thresholds can
be defined to achieve a given efficiency for identifying electrons as
such. The performance can be evaluated by extracting the resulting
pion rejection. Fig.~\ref{fig:pion_rej_mc} shows the performance for
the one-dimensional method from a study using the tracklet simulation
discussed in the next section~\cite{th:hess}.

\section{Simulation}
\label{sec:sim}

\begin{figure}
  \centering
  \includegraphics[width=0.4\textwidth]{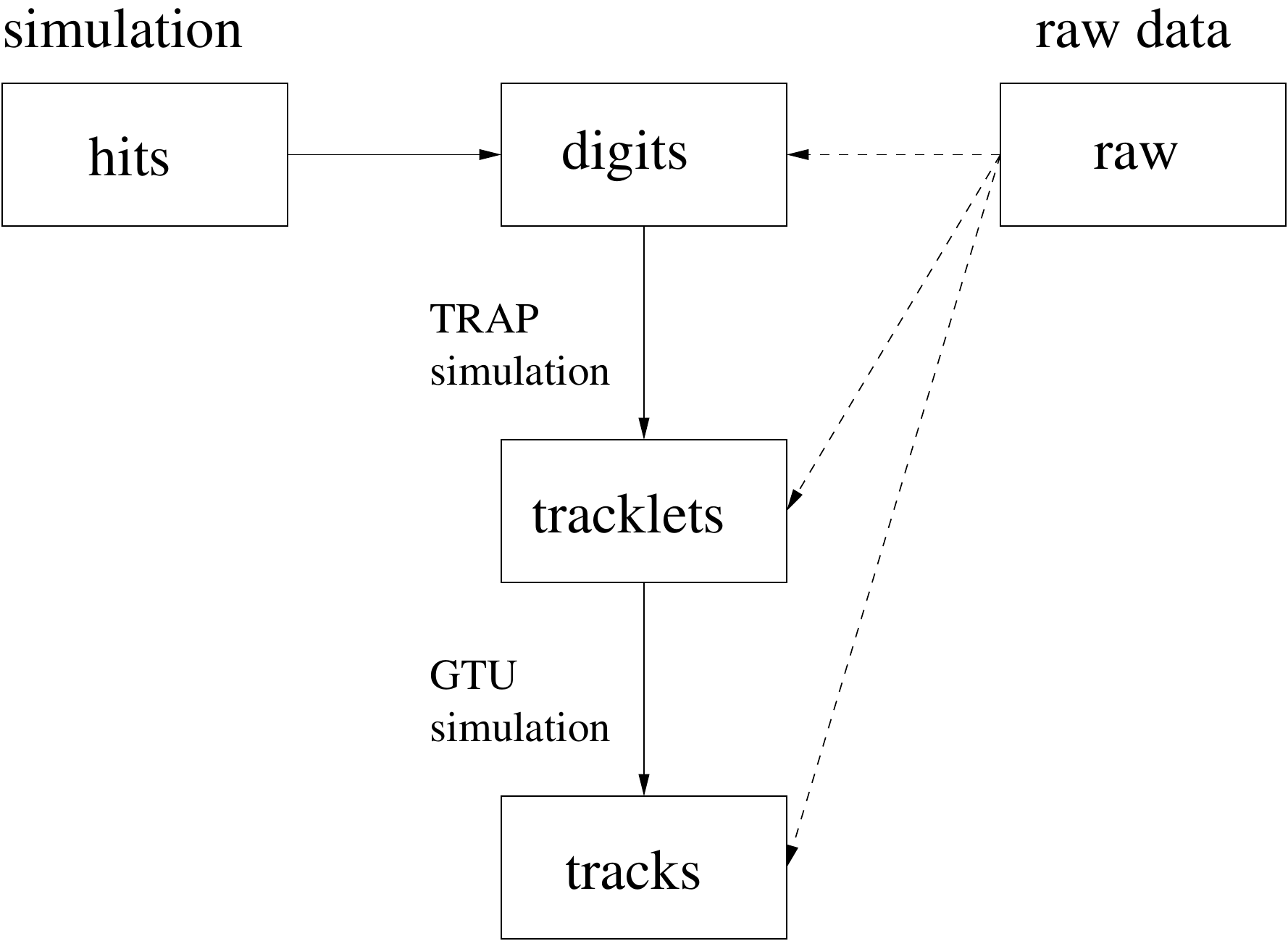}
  \caption{Trigger information in raw data and simulation. During
    reconstruction the ADC values, tracklets and tracks are extracted
    from raw data. The TRAP and GTU simulations allow to recalculate
    the tracklets and tracks from these data. In the pure Monte-Carlo
    case the ADC values (digits) are derived from detector hits.}
  \label{fig:trgsim}
\end{figure}

With the complex processing in multiple stages of the detector it was
crucial for debugging and understanding to have an exact simulation of
the hardware components. For both the local and the global tracking,
models have been implemented in AliRoot~\cite{code:aliroot,th:jkl}
which simulate the behaviour of the electronics by performing exactly
the same operations. Not only can this simulation be used in pure
Monte-Carlo events but also on recorded data. This is because the
trigger is completely based on ADC data which are also read out for
accepted events. In Fig.~\ref{fig:trgsim} the different types of data
are shown. In addition to the ADC values (digits) the raw data contain
the tracklets and the tracks as calculated in the detector electronics
and used for triggering. The simulation can be used to recalculate and
compare them to the actual results.

Also the configuration of the TRAP with all registers and memory
blocks is exactly modelled in AliRoot such that a configuration from
the Detector Control System~\cite{proc:trd:dcs} can be used
identically. This allows to perform simulations with the exact
configuration as used in the actual data taking.

It has been very useful for debugging and understanding to have this
simulation chain available, which agrees on the bit-level with the
data read out from the actual detector. With identical configurations
any deviation between simulation and real data points to a problem. On
the other hand, the effect of parameter changes can be studied much
quicker in the simulation.

\section{Tracking performance}
\label{sec:perf}

\subsection{Monte-Carlo}

Using the simulation of local and global tracking explained in the
previous section it is possible to evaluate the performance based on
events from pure Monte-Carlo generators. The tracklets can be assigned
to known particle trajectories by labels such that position and
deflection can be compared to Monte-Carlo truth. This has mainly been
used to understand the influence of configuration changes, e.g. the
impact on the electron identification.

\subsection{Refit to data}

The tracking results in the raw data do not only contain the final
parametrization but also the references to the contributing
tracklets. Hereby, it is possible to perform a helix fit through the
contributing tracklets and calculate the residuals. Since no other
data but TRD trigger information are needed this can be used as a
quality measure for on-line monitoring.

\subsection{Comparison to off-line tracking}

\begin{figure}
  \centering
  \includegraphics[width=0.4\textwidth]{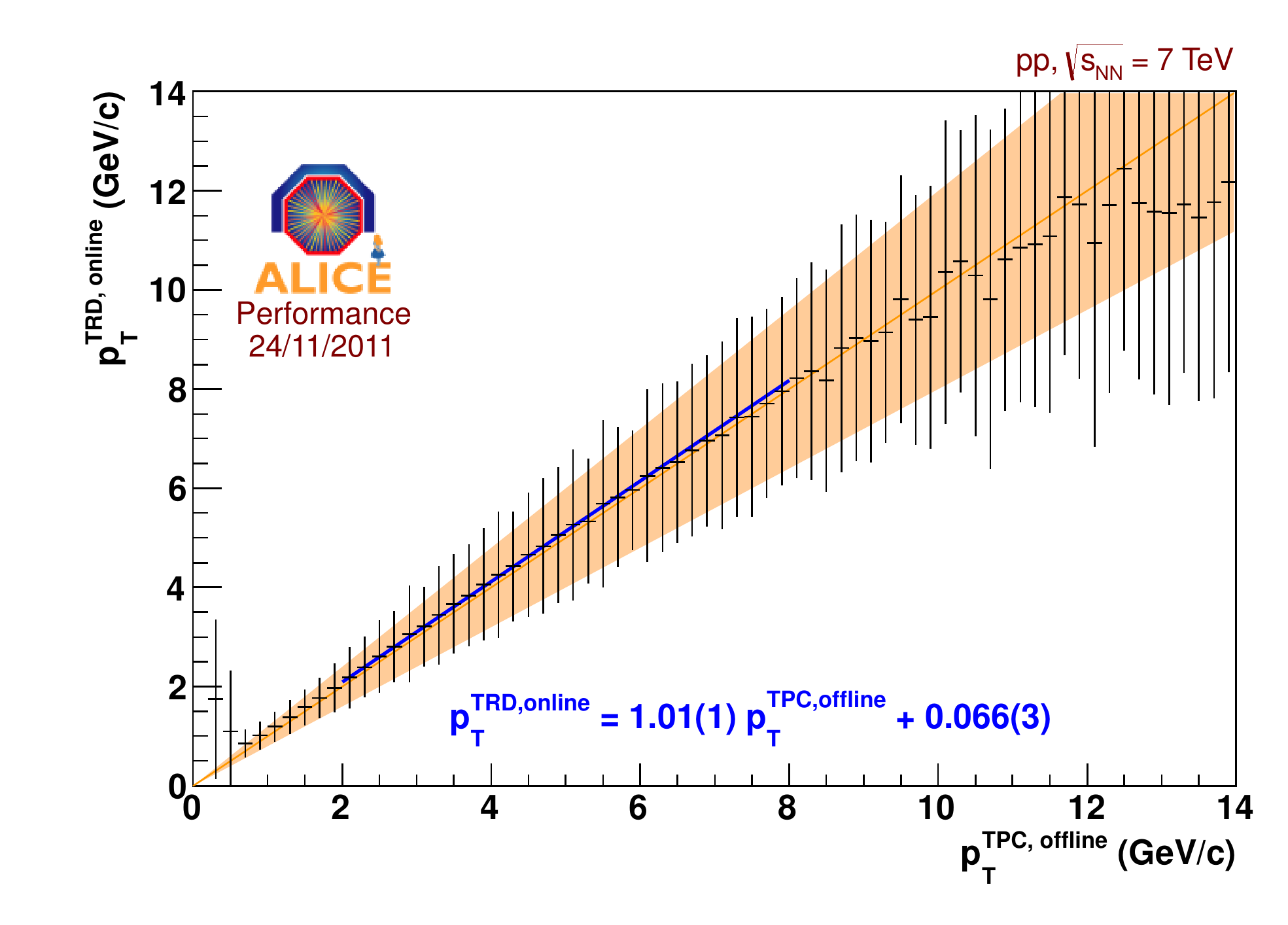}
  \caption{The correlation of the transverse momenta is shown for
    on-line tracks matched to a TPC track. The data point is the mean
    of the distribution and the error bars show the standard
    deviation.}
  \label{fig:pt_corr}
\end{figure}

When the full off-line information is available, tracks reconstructed
in the time projection chamber can be used as reference. They are
geometrically matched to the GTU tracks by the position of
contributing tracklets. Then, the parameters obtained from the on-line
reconstruction can be compared to the off-line reconstruction. This is
shown for the transverse momentum in Fig.~\ref{fig:pt_corr} based on
recently recorded data. As expected the transverse momenta from
on-line and off-line reconstruction are correlated. With 20~\% the
spread is still larger than the design goal. This is due to
non-optimal settings in the front-end electronics, e.g. mis-alignment
is not yet corrected for.

In a similar way the overall tracking efficiency is determined. The
denominator is formed by tracks having at least four tracklets in one
TRD stack. Tracks found in the GTU and matched to a findable one are
considered for the efficiency calculation. This is done to separate
the trigger-specific efficiency from other effects such as the
detector acceptance. For tracks above $3~\mathrm{GeV/c}$ this
efficiency is found to be above 90~\%.

\section{Jet trigger}
\label{sec:jet_trigger}

The area covered by a TRD stack in the $\eta$-$\varphi$ plane is
comparable to the area of a typical jet cone. Therefore, a jet can be
characterized by a number of tracks within such an area and a jet
trigger can be realized by requiring a minimum number of tracks above
a \pt threshold within any TRD stack. This works despite the fact that
only charged tracks are amenable to the TRD reconstruction. From
Monte-Carlo checks we expect the trigger to become fully efficient
from jet energies around $100~\mathrm{GeV}/c$ on, depending on the
condition used.

\begin{figure}
  \centering
  \includegraphics[width=0.4\textwidth]{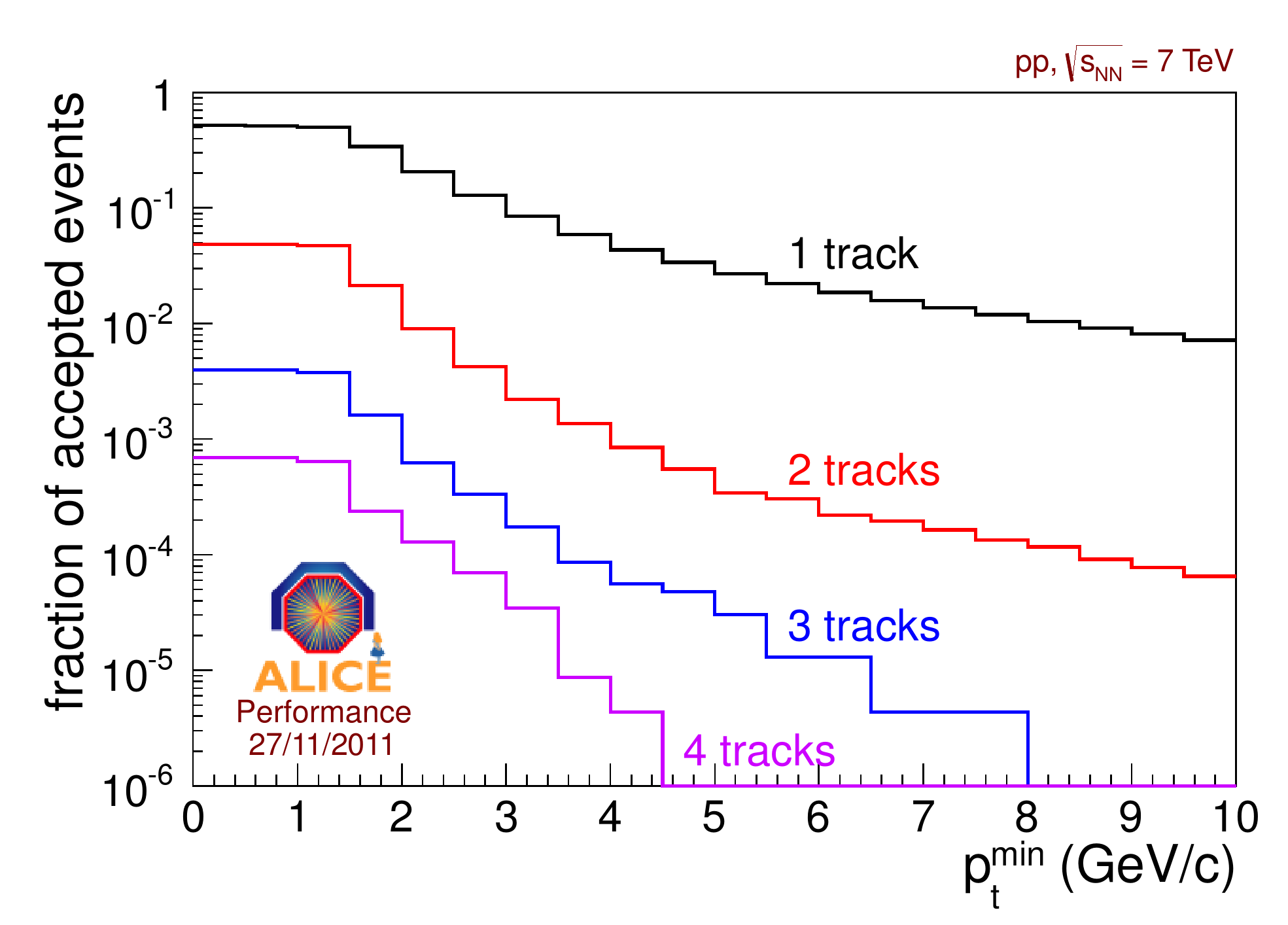}
  \caption{The rejection of minimum bias events is extracted for
    different thresholds in the jet trigger condition. The different
    lines correspond to different minimum numbers of tracks whereas
    the abscissa shows the required \pt.}
  \label{fig:mb_rej}
\end{figure}

Requiring $3$~tracks above $3\, \mathrm{GeV/c}$ in any TRD stack was
found to be a suitable condition to trigger on
jets. Fig.~\ref{fig:mb_rej} shows the rejection of minimum bias events
for different combinations of the minimum number of tracks and the
applied \pt threshold. The various conditions were evaluated on the
tracks in raw data. The above threshold results in a good rejection of
about $10^4$.

\begin{figure}
  \centering
  \includegraphics[width=0.4\textwidth]{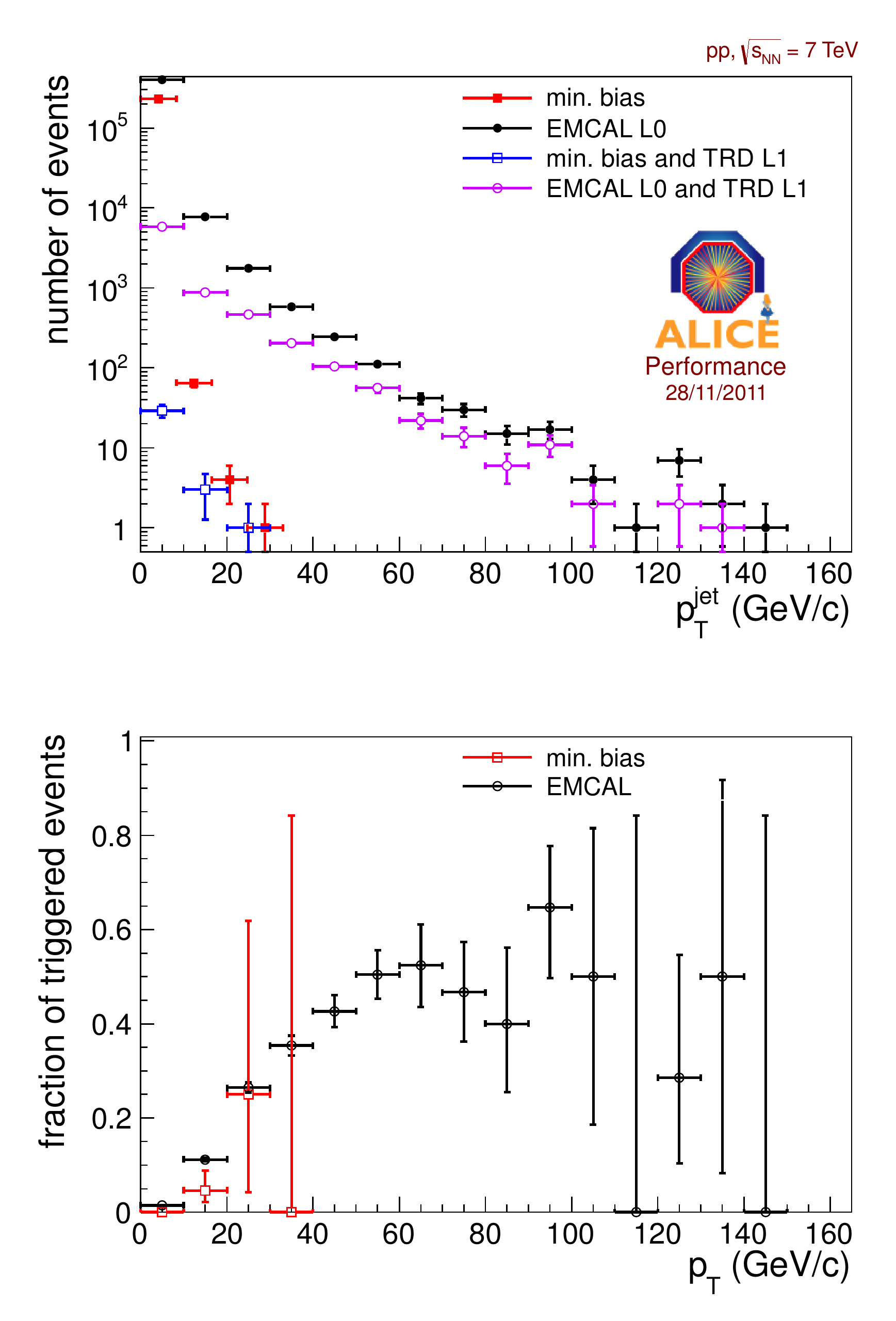}
  \caption{Top: The raw jet spectra are shown for a minimum bias and
    an EMCAL-triggered event sample. In both cases also the spectrum
    for which the trigger condition was fulfilled is plotted. Bottom:
    The efficiency of the jet trigger can be extracted from comparison
    of the triggered and untriggered spectra.}
  \label{fig:jet_spectra}
  \label{fig:jet_eff}
\end{figure}

To check that the jet trigger works as expected, a sample of pp
collisions at $\sqrt{s} = 7\, \mathrm{TeV}$ was used. The events were
classified according to the leading jet with $|\eta| < 0.5$ as
obtained from a UA1 jet finder with $R \simeq 0.4$. The spectra of
triggered events and all events are shown in
Fig.~\ref{fig:jet_eff}. Due to limited statistics in the minimum bias
sample (only down-scaled trigger) also events selected by a tower
trigger from the EMCAL are shown. However, in the current installation
only half of the EMCAL is covered by TRD acceptance. The ratio of the
triggered and untriggered sample shown in Fig.~\ref{fig:jet_eff} must
not be interpreted as a trigger efficiency since there is only a
partial geometrical overlap of EMCAL and TRD at the current
installation status.

\section{Conclusions and Outlook}

Based on first experience in pp data taking we have shown that the
ALICE TRD is well suited to contribute physics triggers to the
experiment at the level-1 stage. The main challenges lie in the very
tight timing constraints, which require strict cuts already on the
tracklet level. Furthermore, all calibration has to be applied on-line
in the front-end electronics. This includes constant geometric
corrections but also quantities varying with ambient conditions, such
as the drift velocity which needs to be stabilized by an active
feedback loop.

Further super-modules will be installed during the winter shutdown and
increase the acceptance. Making use of the electron identification for
the trigger contribution looks promising for future use.

\bibliographystyle{elsarticle-num}
\bibliography{ref}

\end{document}

%% file: fig/trd_trigger.pdftex_t
\begin{picture}(0,0)%
\includegraphics{fig/trd_trigger.pdf}%
\end{picture}%
\setlength{\unitlength}{4144sp}%
\begingroup\makeatletter\ifx\SetFigFont\undefined%
\gdef\SetFigFont#1#2#3#4#5{%
  \reset@font\fontsize{#1}{#2pt}%
  \fontfamily{#3}\fontseries{#4}\fontshape{#5}%
  \selectfont}%
\fi\endgroup%
\begin{picture}(10207,4860)(-734,-4351)
\put(3354,-3031){\makebox(0,0)[b]{\smash{{\SetFigFont{11}{13.2}{\familydefault}{\mddefault}{\updefault}{\color[rgb]{0,0,0}Global Tracking}%
}}}}
\put(-179,-736){\rotatebox{90.0}{\makebox(0,0)[b]{\smash{{\SetFigFont{20}{24.0}{\familydefault}{\mddefault}{\updefault}{\color[rgb]{0,0,0}Electronics}%
}}}}}
\put(-179,-3076){\rotatebox{90.0}{\makebox(0,0)[b]{\smash{{\SetFigFont{20}{24.0}{\familydefault}{\mddefault}{\updefault}{\color[rgb]{0,0,0}Tracking Unit}%
}}}}}
\put(-494,-736){\rotatebox{90.0}{\makebox(0,0)[b]{\smash{{\SetFigFont{20}{24.0}{\familydefault}{\mddefault}{\updefault}{\color[rgb]{0,0,0}Front End}%
}}}}}
\put(-494,-3076){\rotatebox{90.0}{\makebox(0,0)[b]{\smash{{\SetFigFont{20}{24.0}{\familydefault}{\mddefault}{\updefault}{\color[rgb]{0,0,0}Global}%
}}}}}
\put(136,254){\makebox(0,0)[lb]{\smash{{\SetFigFont{20}{24.0}{\familydefault}{\mddefault}{\updefault}{\color[rgb]{0,0,0}Pretrigger}%
}}}}
\put(4413,119){\makebox(0,0)[lb]{\smash{{\SetFigFont{20}{24.0}{\familydefault}{\mddefault}{\updefault}{\color[rgb]{0,0,0}L1 arrival}%
}}}}
\put(856,-331){\makebox(0,0)[lb]{\smash{{\SetFigFont{20}{24.0}{\familydefault}{\mddefault}{\updefault}{\color[rgb]{0,0,0}L0 arrival}%
}}}}
\put(7246,-2131){\makebox(0,0)[b]{\smash{{\SetFigFont{20}{24.0}{\familydefault}{\mddefault}{\updefault}{\color[rgb]{0,0,0}80-500}%
}}}}
\put(676,-2221){\makebox(0,0)[b]{\smash{{\SetFigFont{20}{24.0}{\familydefault}{\mddefault}{\updefault}{\color[rgb]{0,0,0}1}%
}}}}
\put(1351,-2221){\makebox(0,0)[b]{\smash{{\SetFigFont{20}{24.0}{\familydefault}{\mddefault}{\updefault}{\color[rgb]{0,0,0}2}%
}}}}
\put(2026,-2221){\makebox(0,0)[b]{\smash{{\SetFigFont{20}{24.0}{\familydefault}{\mddefault}{\updefault}{\color[rgb]{0,0,0}3}%
}}}}
\put(2656,-2221){\makebox(0,0)[b]{\smash{{\SetFigFont{20}{24.0}{\familydefault}{\mddefault}{\updefault}{\color[rgb]{0,0,0}4}%
}}}}
\put(3376,-2221){\makebox(0,0)[b]{\smash{{\SetFigFont{20}{24.0}{\familydefault}{\mddefault}{\updefault}{\color[rgb]{0,0,0}5}%
}}}}
\put(4051,-2221){\makebox(0,0)[b]{\smash{{\SetFigFont{20}{24.0}{\familydefault}{\mddefault}{\updefault}{\color[rgb]{0,0,0}6}%
}}}}
\put(4726,-2221){\makebox(0,0)[b]{\smash{{\SetFigFont{20}{24.0}{\familydefault}{\mddefault}{\updefault}{\color[rgb]{0,0,0}7}%
}}}}
\put(9361,-2131){\makebox(0,0)[rb]{\smash{{\SetFigFont{20}{24.0}{\familydefault}{\mddefault}{\updefault}{\color[rgb]{0,0,0}$\mu$s}%
}}}}
\put(4141,-4336){\makebox(0,0)[lb]{\smash{{\SetFigFont{20}{24.0}{\familydefault}{\mddefault}{\updefault}{\color[rgb]{0,0,0}Contribution at CTP}%
}}}}
\put(4411,-4021){\makebox(0,0)[lb]{\smash{{\SetFigFont{20}{24.0}{\familydefault}{\mddefault}{\updefault}{\color[rgb]{0,0,0}L1 arrival}%
}}}}
\put(7291,-4021){\makebox(0,0)[lb]{\smash{{\SetFigFont{20}{24.0}{\familydefault}{\mddefault}{\updefault}{\color[rgb]{0,0,0}L2a arrival}%
}}}}
\put(856,-4021){\makebox(0,0)[lb]{\smash{{\SetFigFont{20}{24.0}{\familydefault}{\mddefault}{\updefault}{\color[rgb]{0,0,0}L0 arrival}%
}}}}
\put(8258,-3346){\makebox(0,0)[b]{\smash{{\SetFigFont{12}{14.4}{\familydefault}{\mddefault}{\updefault}{\color[rgb]{0,0,0}Data transfer to DAQ}%
}}}}
\put(5613,-2448){\makebox(0,0)[b]{\smash{{\SetFigFont{12}{14.4}{\familydefault}{\mddefault}{\updefault}{\color[rgb]{0,0,0}Receive raw data}%
}}}}
\put(5613,-1361){\makebox(0,0)[b]{\smash{{\SetFigFont{12}{14.4}{\familydefault}{\mddefault}{\updefault}{\color[rgb]{0,0,0}Raw data shipping to GTU}%
}}}}
\put(676,-736){\makebox(0,0)[b]{\smash{{\SetFigFont{12}{14.4}{\familydefault}{\mddefault}{\updefault}{\color[rgb]{0,0,0}Drift time}%
}}}}
\put(788,-1058){\makebox(0,0)[b]{\smash{{\SetFigFont{12}{14.4}{\familydefault}{\mddefault}{\updefault}{\color[rgb]{0,0,0}Fit calculation}%
}}}}
\put(1981,-1321){\makebox(0,0)[b]{\smash{{\SetFigFont{12}{14.4}{\familydefault}{\mddefault}{\updefault}{\color[rgb]{0,0,0}building}%
}}}}
\put(1981,-1051){\makebox(0,0)[b]{\smash{{\SetFigFont{12}{14.4}{\familydefault}{\mddefault}{\updefault}{\color[rgb]{0,0,0}Tracklet}%
}}}}
\put(2476,-1636){\makebox(0,0)[lb]{\smash{{\SetFigFont{12}{14.4}{\familydefault}{\mddefault}{\updefault}{\color[rgb]{0,0,0}Tracklet shipping}%
}}}}
\put(2746,-2716){\makebox(0,0)[lb]{\smash{{\SetFigFont{12}{14.4}{\familydefault}{\mddefault}{\updefault}{\color[rgb]{0,0,0}Receive tracklets}%
}}}}
\put(3961,-3616){\makebox(0,0)[lb]{\smash{{\SetFigFont{12}{14.4}{\familydefault}{\mddefault}{\updefault}{\color[rgb]{0,0,0}Ship contribution}%
}}}}
\end{picture}%